\documentclass[twoside]{article}
\usepackage{adjustbox}
\usepackage[accepted]{aistats2025}
\usepackage{xcolor}
\usepackage{xurl}

\newcommand{\sG}{\mathsf{G}}
\newcommand{\sV}{\mathsf{V}}
\newcommand{\sE}{\mathsf{E}}

\newcommand{\prob}[1]{\mathbb{P}\left[#1\right]}

\usepackage{amsfonts,amsmath,amsthm,subcaption}

\newtheorem{theorem}{Theorem}[section]
\newtheorem{lemma}[theorem]{Lemma}
\newtheorem{definition}{Definition}

\usepackage[linesnumbered,ruled,vlined]{algorithm2e}
\begin{document}

%

%

\twocolumn[

\aistatstitle{Counting Graphlets of Size $k$ under Local Differential Privacy}

\aistatsauthor{ Vorapong Suppakitpaisarn\footnotemark[1] \And Donlapark Ponnoprat\footnotemark[1] \And  Nicha Hirankarn \And Quentin Hillebrand }

\aistatsaddress{The University of Tokyo \And Chiang Mai University    \And Tsukuba University \And The University of Tokyo } ]

\footnotetext[1]{Equal contribution.}

\begin{abstract}
The problem of counting subgraphs or graphlets under local differential privacy is an important challenge that has attracted significant attention from researchers. However, much of the existing work focuses on small graphlets like triangles or \( k \)-stars. In this paper, we propose a non-interactive, locally differentially private algorithm capable of counting graphlets of any size \( k \). When $n$ is the number of nodes in the input graph, we show that the expected \( \ell_2 \) error of our algorithm is \( O(n^{k - 1}) \). Additionally, we prove that there exists a class of input graphs and graphlets of size \( k \) for which any non-interactive counting algorithm incurs an expected \( \ell_2 \) error of \( \Omega(n^{k - 1}) \), demonstrating the optimality of our result. Furthermore, we establish that for certain input graphs and graphlets, any locally differentially private algorithm must have an expected \( \ell_2 \) error of \( \Omega(n^{k - 1.5}) \). Our experimental results show that our algorithm is more accurate than the classical randomized response method.
\end{abstract}

\section{INTRODUCTION}

In recent years, differential privacy~\cite{dwork2006differential,dwork2014algorithmic} has gained recognition as the leading approach for ensuring strong privacy protections while still enabling effective data analysis. It achieves this by ensuring that the results of computations remain nearly the same, even if the data of a single individual is altered, thereby protecting personal information. While the initial focus of differential privacy was on tabular datasets~\cite{dwork2006calibrating,mcsherry2007mechanism}, there has been a growing interest in extending these privacy protections to graph data~\cite{sajadmanesh2021locally,ye2020lf}, which presents a unique set of complexities and challenges.

Differential privacy has diversified into several variants to address different use cases, as outlined in \cite{desfontaines2019sok}. Of particular relevance to our work is local differential privacy~\cite{cormode2018privacy,evfimievski2003limiting}, which differs from traditional approaches by eliminating the need for a trusted central server. Instead, users anonymize their own private data before transmitting it to an untrusted party. In the realm of graph data, the most widely used variant is edge local differential privacy~\cite{qin2017generating}, where each user's sensitive data relates to their relationships or connections with others in the graph.

A frequently used obfuscation method is the randomized response technique \cite{warner_randomized_1965}. In this method, users randomly flip bits in their adjacency vector with a specified probability. The server then aggregates this modified data to create an obfuscated version of the graph. While various graph statistics can be derived from this obfuscated graph, the accuracy of the information is often limited. Algorithms tailored to publish specific statistics generally provide more precise and valuable insights into the graph.

A commonly studied graph statistic in the context of local differential privacy is the number of subgraphs or graphlets \cite{imola2021locally,hillebrand2023communication}. 
Many studies have focused specifically on publishing triangle counts  \cite{imola2022communication,hillebrand2023communication,eden2023triangle,liu2024edge}. For example, \cite{imola2021locally} introduces a non-interactive, edge-local differentially private algorithm. When $n$ is the number of nodes in the input graph, the authors show that the expected \(\ell_2\)-error of the counting algorithm is \(O(n^2)\). On the other hand, \cite{eden2023triangle} proves that any non-interactive, edge-local differentially private algorithm will have an \(\ell_2\)-error of \(\Omega(n^2)\). Additionally, \cite{eden2023triangle} further demonstrates that any edge-local differentially private algorithm would result in an \(\ell_2\)-error of \(\Omega(n^{1.5})\).

Counting graphlets beyond triangles plays a crucial role in various applications \cite{ahmed2015efficient,bressan2017counting,marcus2010efficient}. For example, in sociometric studies, the profile of each subgraph of size \(k\) is typically referred to as a \(k\)-subgraph census. This census is employed in various social network analyses \cite{holland1976local}. In computer networks, the number of bi-fan motifs can reflect the frequency of points of presence \cite{feldman2008automatic}. \textcolor{black}{Those applications often rely on social network data, where node connections may contain sensitive information. Users may prefer not to share their connection details with central servers, requiring local obfuscation of the data.} 

This motivates us to address this problem within the edge local differential privacy framework in this paper. However, to the best of our knowledge, only works address the counting of larger subgraphs are: \cite{hillebrand2024}, which presents an algorithm for counting odd-length cycles, \cite{he2024butterfly}, which counts the number of cycles with length four in bipartite graphs, \cite{imola2022differentially}, which counts the number of cycles with length four but under a weaker assumption called as shuffle model, and \cite{betzerpublishing}, which counts walks of any length.

\subsection{Our Contributions}

Our contributions can be summarized as follows:
\begin{enumerate}
    \item In Section 3, we give a non-interactive and edge-local differentially private algorithm for counting any graphlets of $k$ nodes for any $k$.
    \item Also, in Section 3, we demonstrate that our counting algorithm has an expected $\ell_2$-error of $O(n^{k - 1})$.
    \item In Section 4, we demonstrate that there is a class of input graphs and a graphlet of $k$ nodes such that \textit{any} non-interactive and edge-local differentially private algorithm would have an expected $\ell_2$-error of $\Theta(n^{k - 1})$. Hence, our algorithm is asymptotically optimal in terms of the expected $\ell_2$-error.
    \item In Section 5, we demonstrate that there is a class of input graphs and a graphlet of $k$ nodes such that \emph{any} edge-local differentlly private algorithm would have an expected $\ell_2$-error of $\Theta(n^{k - 1.5})$.
\end{enumerate}

\begin{table}[h!]
\centering
\begin{tabular}{|l|l|}
\hline
\textbf{Graphlet Type} & \textbf{Results} \\ \hline
Triangle               & $O(n^2)$ (non-interactive) \\ 
                       & \cite{imola2021locally} \\
                       & $\Omega(n^2)$ (non-interactive)  \\ 
                       &  \cite{eden2023triangle}\\
                       & $\Omega(n^{1.5})$ (any algorithm)  \\ 
                       &  \cite{eden2023triangle}\\ \hline
Graphlet  & $O(n^{k-1})$ (non-interactive)  \\ 
  with $k$ nodes                     &  [Section 3] \\
                       & $\Omega(n^{k-1})$ (non-interactive)  \\ 
                       & [Section 4] \\
                       & $\Omega(n^{k-1.5})$ (any algorithm)  \\ 
                       & [Section 5] \\ \hline
\end{tabular}
\caption{Our results in this paper compared with the previous results}
\label{tab1}
\end{table}

We summarize our results in comparison to previous works in Table \ref{tab1}. For the case where the graphlet is a triangle, we set $k = 3$. Our findings are consistent with those of prior research \cite{imola2021locally,eden2023triangle}, and can be viewed as a generalization of the results on triangle counting. While we draw on some of the algorithmic and proof techniques from these earlier studies, our results are significantly more general. Consequently, we had to introduce several additional proof methods, making our proofs more complex and involved.

While we believe that the algorithm proposed in Section 3 is valuable for local differential privacy research, the lower bound results presented in Sections 4 and 5 are equally significant. These results demonstrate that the expected $\ell_2$-error can increase by a factor of $n$ when the size of the graphlets increases by one. This suggests that obtaining precise results for large graphlet counting under local differential privacy on general graphs may not be possible. Therefore, developing algorithms tailored to specific types of graphs, as done in \cite{hillebrand2024,betzerpublishing}, is crucial.

In addition to the theoretical contributions, Section 6 presents experiments that validate the effectiveness of our algorithm. Since no existing algorithm is specifically designed for subgraph counting, we compare our approach with the classical randomized response technique. Our algorithm achieves an $\ell_2$-error up to 36 times smaller than that of randomized response when $n = 100$. Moreover, we observe that the performance gap increases as $n$ grows, suggesting that the improvement becomes significantly more pronounced for larger graphs.

\subsection{Related Works}

Graph data mining under local differential privacy is still an emerging area, whereas differential privacy has been explored for many years in various studies, including \cite{gupta2010differentially,olatunji2021releasing}. According to \cite{imola2021locally}, local differential privacy generally focuses on protecting the privacy of edges or relationships, with a few exceptions such as \cite{zhang2020differentially}. In contrast, differential privacy at large can conceal whether a specific individual or node belongs to a social network, as seen in \cite{hay2009accurate,raskhodnikova2016differentially}. While both edge and node privacy models exist, node-level privacy is not applicable within the scope of local differential privacy.

In addition to the subgraph counting, there are many graph algorithms proposed in the framework of local differential privacy. For examples, recent research has developed techniques for estimating the densest subgraph, \(k\)-core decomposition, and degeneracy within the local differential privacy framework \cite{dhulipala2022differential, dhulipala2023near, dinitz2023improved, henzinger2024tighter}. 

\section{PRELIMINARIES}

\subsection{Notations}

Let the input graph be \( G = (V, E) \), where \( V = \{v_1, \dots, v_n\} \) and \( E \subseteq V^2 \). Each node \( v_i \) represents a user in our social network, and the edges \( E \) represent the relationships between them. Now, consider a graphlet \( \sG = (\sV, \sE) \) with \( \sV = \{u_1, \dots, u_k\} \) and $\sE \subseteq \sV^2$. The number of graphlets \(\mathsf{G}\) in \( G \), denoted as \( \mathsf{G}(G) \), is defined as the number of distinct subgraphs \( (V', E') \subseteq G \) that are isomorphic to \( \mathsf{G} \).

For each \( i \in [1,n] \), let \( a_i = [a_{i,1}, \ldots, a_{i,n}] \) be the adjacency list of user \( v_i \), where \( a_{i,j} = 1 \) indicates an edge between \( v_i \) and \( v_j \) (i.e., \( (v_i, v_j) \in E \)), and \( a_{i,j} = 0 \) otherwise. The degree of node \( v_i \), denoted by \( d_i \), represents the number of edges connected to \( v_i \). At the start of any locally differentially private algorithm, we assume that each user \( v_i \) has knowledge of only their own adjacency vector \( a_i \). The vector is a sensitive information of the user.

\textcolor{black}{We deploy the expected $\ell_2$-error to measure the precision of the graphlet counting algorithms in this work. Let $\tilde{\mathsf{G}}(G)$ be the estimation of the number of graphlets obtaind from the algorithm. The expected $\ell_2$-error is defined as:
\[
\sqrt{\int_{-\infty}^{\infty} \Pr[\tilde{\mathsf{G}}(G) = i] (\mathsf{G}(G) - i)^2 \, di}.
\]
This expected 
$\ell_2$-error is widely recognized as a standard approach for evaluating the precision of various non-deterministic algorithms such as \cite{drews2023analysis,huang2024making}.} 

\subsection{Edge Local Differential Privacy}

We define two adjacency lists, \(a\) and \(a'\), as neighbors if they differ by exactly one bit, meaning that one can be transformed into the other by adding or removing a single edge involving node \( v_i \). 
The concept of edge local differential privacy can be described as follows:

\begin{definition}[Local differentially private query] 
\label{ref1} Let \(\epsilon > 0\). A randomized query \(\mathcal{R}\) is said to be \(\epsilon\)-edge locally differentially private for node \(v_i\) if, for any pair of neighboring adjacency lists \(a\) and \(a'\), and for any possible set of outcomes \(S\), we have that
    $\prob{\mathcal{R}(a) \in S} \leq e^{\epsilon} \prob{\mathcal{R}(a') \in S}$.

\end{definition}

\begin{definition}[\cite{qin2017generating}]
\label{dif:diff}
An algorithm \(\mathcal{A}\) is defined as \(\epsilon\)-edge locally differentially private if, for any user $v_i$ and for any possible set of queries \(\mathcal{R}_1, \dots, \mathcal{R}_\kappa\) which $\mathcal{A}$ posed to user \(v_i\), where each query \(\mathcal{R}_j\) is \(\epsilon_j\)-edge locally differentially private (for \(1 \leq j \leq \kappa\)), the total privacy loss satisfies \(\epsilon_1 + \cdots + \epsilon_\kappa \leq \epsilon\).
\end{definition}
When an algorithm A is $\epsilon$-edge locally differentially private, it is also said to have a privacy budget of $\epsilon$.
In the subsequent definition, we define the non-interactive algorithm.
\begin{definition}[Non-interactive algorithm] An algorithm \(\mathcal{A}\) is considered non-interactive if it issues a single query to all users, and the query is independent of the responses from other users.
\end{definition}

\subsection{Unbiased Randomized Response}
In this subsection, we examine the randomized response query, designed to release a privacy-preserving version of the adjacency vector.

\begin{definition}[\cite{warner_randomized_1965,wang2016using}]
For \(\epsilon > 0\), the randomized response query $\mathcal{R}$ with privacy budget $\epsilon$ operates on an adjacency list \(a = (a_1, \dots, a_n)\), producing an obfuscated output \(\tilde{a} = (\tilde{a}_1, \dots, \tilde{a}_n)\) such that
\[
\prob{\mathcal{R}(\tilde{a}_i) = 1} = 
\begin{cases}
    \frac{e^{\epsilon}}{1 + e^{\epsilon}} & \text{if } a_i = 1, \\
    \frac{1}{1 + e^{\epsilon}} & \text{if } a_i = 0.
\end{cases}
\]
This query guarantees \(\epsilon\)-edge local differential privacy.
\end{definition}
A graph \(\tilde{G}\) can be constructed using the collection of obfuscated adjacency vectors from all users. By analyzing the statistics of this obfuscated graph \(\tilde{G}\), we can release various insights, such as the number of subgraphs \cite{ye2020towards,imola2021locally,imola2022communication,hillebrand2023communication}. In Section 6, we will use the estimates derived from \(\tilde{G}\) as our benchmark for comparison. Also, we denote the obfuscated version of variable $a_{i,j}$ as $\tilde{a}_{i,j}$, i.e. $\tilde{a}_{i,j} = \mathcal{R}(a_{i,j})$ when $\mathcal{R}$ is the randomized response query.

The variable \(\tilde{a}_{i,j}\) provides a biased estimate of \(a_{i,j}\), as \(\mathbb{E}[\tilde{a}_{i,j}] \neq a_{i,j}\). In the triangle counting algorithm proposed by \cite{eden2023triangle}, the authors address this bias by using an adjusted estimator \(\hat{a}_{i,j} := \frac{e^{\epsilon} + 1}{e^{\epsilon} - 1} \tilde{a}_{i,j} - \frac{1}{e^{\epsilon} - 1}\) to estimate the number of triangles. They demonstrate that this adjustment ensures \(\mathbb{E}[\hat{a}_{i,j}] = a_{i,j}\). 

\section{OUR ALGORITHM}

Algorithm \ref{alg1} presents the method for estimating the number of graphlets \(\sG = (\mathsf{V} = \{u_1, \dots, u_k\}, \mathsf{E})\) in \(G = (V = \{v_1, \dots, v_n\}, E)\) under local differential privacy. A bijective function \( \pi: \mathsf{V} \rightarrow \mathsf{V} \) is an automorphism of \( \sG \) if \( (\pi(u_i), \pi(u_j)) \in \mathsf{E} \) if and only if \( (u_i, u_j) \in \mathsf{E} \). In the algorithm, we use \( A(\sG) \) to denote number of automorphisms of \( \sG \). At Line 1, we apply the randomized response technique to collect edge information from all users. In Line 2, we adjust the results to remove bias introduced by the randomized response. Finally, in Lines 3-4, we compute the number of graphlets based on the unbiased results.

\begin{algorithm}
    \caption{Estimate the number of graphlets under local differential privacy}
    \label{alg1}
        \KwIn{Graph $G = (V,E)$, privacy budget $\epsilon$, graphlet $\sG = (\sV, \sE)$, number of automorphisms of $\sG$ denoted by $A(\sG)$}
        \KwOut{Estimation of the number of graphlets \(\sG\) in \(G\)}
        \textbf{[All users and server]} Inquire the  randomized response query with privacy
budget $\epsilon$ to all users. Let the result be $\tilde{a}_{i,j}$ for all $i,j$.\\
        \textbf{[Server]} For all $i,j$, calculate $\hat{a}_{i,j} := \frac{e^{\epsilon} + 1}{e^{\epsilon} - 1} \tilde{a}_{i,j} - \frac{1}{e^{\epsilon} - 1}$. \\
        \textbf{[Server]}  Let $\mathcal{D}$ be the set of tuples \( \mathcal{W} = (v_{\ell_1}, \dots, v_{\ell_k}) \in V^k \) where $\ell_i \in [n]$ such that \( v_{\ell_i} \neq v_{\ell_j} \) for all \( i \neq j \). For all $\mathcal{W} \in \mathcal{D}$, calculate $\tilde{W}(\mathcal{W}, \sG) = \prod\limits_{\{u_i,u_j\} \in \mathsf{E}} \hat{a}_{\ell_i,\ell_j}$.\\
        \textbf{[Server]} \Return $\tilde{\sG}(G) = \left(\sum\limits_{\mathcal{W} \in \mathcal{D}} \tilde{W}(\mathcal{W}, \sG) \right) / A(\sG)$. \\
\end{algorithm}

It is clear that the algorithm is non-interactive and satisfies $\epsilon$-edge local differential privacy, since the only query posed to users is the randomized response query in Line 1. In the remaining part of this section, we demonstrate that the $\ell_2$-error of this algorithm is $O(n^{k - 1})$.

 Consider a tuple \( \mathcal{W} = (v_{\ell_1}, \dots, v_{\ell_k}) \in V^k \) where $\ell_i \in [n]$ such that \( v_{\ell_i} \neq v_{\ell_j} \) for all \( i \neq j \), and for all \( \{u_i, u_j\} \in \mathsf{E} \), the edge \( \{v_{\ell_i}, v_{\ell_j}\} \) is in \( E \). Let \( W(G, \mathsf{G}) \) be the number of such tuples. We obtain the following lemma.

 \begin{lemma}
\(\mathsf{G}(G)= W(G, \mathsf{G}) / A(\sG) \). \label{lem1}
\end{lemma}
\begin{proof}
Let \(\mathsf{S}\) be the set of subgraphs of \(G\) isomorphic to \(\mathsf{G}\), and let \(\mathsf{W}\) be the set of tuples \(\mathcal{W} = (v_{\ell_1}, \dots, v_{\ell_k}) \in V^k\) such that \(v_{\ell_i} \neq v_{\ell_j}\) for all \(i \neq j\), and for all \(\{u_i, u_j\} \in \mathsf{E}\), the edge \(\{v_{\ell_i}, v_{\ell_j}\}\) is in \(E\). To demonstrate this lemma, we provide a surjective function \(g: \mathsf{W} \rightarrow \mathsf{S}\) such that \(|g^{-1}(S)| = A(\sG)\) for all \(S \in \mathsf{S}\).

For \(\mathcal{W} = (v_{\ell_1}, \dots, v_{\ell_k}) \in \mathsf{W}\), we define a subgraph \(g(\mathcal{W}) = (V', E')\) such that \(V' = \{v_{\ell_1}, \dots, v_{\ell_k}\}\) and \(E' = \{\{v_{\ell_i}, v_{\ell_j}\} : \{u_i, u_j\} \in \mathsf{E}\}\). By the definition of \(\mathsf{W}\), \(g(\mathcal{W})\) is in \(\mathsf{S}\).

On the other hand, consider a subgraph \((V' = \{v_{\ell_1}, \dots, v_{\ell_k}\}, E') \in \mathsf{S}\). We have that a tuple \(\mathcal{W} = (v_{q_1}, \dots, v_{q_k}) \in \mathsf{W}\) has \(g(\mathcal{W}) = (V', E')\) if 1)
    \(\{v_{q_1}, \dots, v_{q_k}\} = \{v_{\ell_1}, \dots, v_{\ell_k}\}\), and 2) \(\{ \{v_{q_i}, v_{q_j}\} : \{u_i, u_j\} \in \mathsf{E} \} = E'\).

Let \(\mathsf{A}(\mathsf{G})\) be the set of automorphisms of \(\mathsf{G}\). For \(\pi \in \mathsf{A}(\mathsf{G})\), define \(\pi': [n] \rightarrow [n]\) such that \(\pi'(i) = j\) if \(\pi(u_i) = u_j\). Because \(\pi\) is an automorphism, we have \(\{u_{\pi'(i)}, u_{\pi'(j)}\} \in \mathsf{E}\) for \(\{u_i, u_j\} \in  \mathsf{E}\). Therefore, \(g(\mathcal{W}) = (V', E')\) if and only if there exists \(\pi \in \mathsf{A}(\sG)\) such that \(v_{\ell_{\pi'(i)}} = v_{q_i}\) for all \(i\). The number of distinct tuples \(\mathcal{W}\) such that \(g(\mathcal{W}) = (V', E')\) is then equal to the number of distinct functions \(\pi\) in \(\mathsf{A}(\mathsf{G})\), which is \(A(\mathsf{G})\).
\end{proof}

We show in the following lemma that the estimated value from our algorithm has no bias.
\begin{lemma}
    The estimator $\tilde{\sG}(G)$ is an unbiased estimator of $\sG(G)$. \label{lem:bias}
\end{lemma}
\begin{proof}
It follows from the independence of $\tilde{a}_{i,j}$'s and Lemma \ref{lem1} that
\begin{eqnarray*}
    & & \mathbb{E}\left[ \tilde{\sG}(G) \right] \\ & = & \mathbb{E}\left[\sum\limits_{(v_{\ell_1}, \dots, v_{\ell_k}) \in \mathcal{D}} \prod\limits_{\{u_i,u_j\} \in \mathsf{E}} \hat{a}_{\ell_i,\ell_j}\right] / A(\mathsf{G}) \\ & = & \sum\limits_{(v_{\ell_1}, \dots, v_{\ell_k}) \in \mathcal{D}} \prod\limits_{\{u_i,u_j\} \in \mathsf{E}}  \mathbb{E} \left[ \hat{a}_{\ell_i,\ell_j} \right] / A(\mathsf{G}) \\
    & = & \sum\limits_{(v_{\ell_1}, \dots, v_{\ell_k}) \in \mathcal{D}} \prod\limits_{\{u_i,u_j\} \in \mathsf{E}}   a_{\ell_i, \ell_j}  / A(\mathsf{G}) \\
    & = & W(G,\sG) / A(\mathsf{G}) = \mathsf{G}(G).
\end{eqnarray*}
\end{proof}

In the following lemma, we consider the property of the random variable \(\tilde{W}(\mathcal{W}, \mathsf{G})\). This property will be used later to determine the variance of our estimator.
\begin{lemma}
There are at most $O(n^{2k - 2})$ pairs of $\mathcal{W}, \mathcal{W}' \in \mathcal{D}$ such that the covariance of $\tilde{W}(\mathcal{W},\sG)$ and $\tilde{W}(\mathcal{W}',\sG)$ is more than 0. \label{lem23}
\end{lemma}
\begin{proof}
Let \(\mathcal{W} = (v_{\ell_1}, \dots, v_{\ell_k}) \in \mathcal{D}\) and \(\mathcal{W}' = (v_{\ell'_1}, \dots, v_{\ell'_k}) \in \mathcal{D}\). We have \(\mathrm{Cov}(\tilde{W}(\mathcal{W}, \sG), \tilde{W}(\mathcal{W}', \sG)) > 0\) only if the calculations of \(\tilde{W}(\mathcal{W}, \sG)\) and \(\tilde{W}(\mathcal{W}', \sG)\) both involve the same randomized data \(\tilde{a}_{\ell_i, \ell_j}\) for some $i,j$. This occurs if \(|\{v_{\ell_1}, \dots, v_{\ell_k}\} \cap \{v_{\ell'_1}, \dots, v_{\ell'_k}\}| \geq 2\). This results in:
\begin{equation}\label{eq:k}
k \leq |\{v_{\ell_1}, \dots, v_{\ell_k}\} \cup \{v_{\ell'_1}, \dots, v_{\ell'_k}\}| \leq 2k - 2.
\end{equation}
Let us denote by $\mathsf{C}(\mathsf{k})$ the number of pairs $\mathcal{W}, \mathcal{W}'$ such that \(|\{v_{\ell_1}, \dots, v_{\ell_k}\} \cup \{v_{\ell'_1}, \dots, v_{\ell'_k}\}| = \mathsf{k}\). We have from~\eqref{eq:k} that $\mathsf{C}(\mathsf{k})>0$ only if $k \leq \mathsf{k} \leq 2k - 2$. There are at most $n^{\mathsf{k}}$ combinations of $\{v_{\ell_1}, \dots, v_{\ell_k}\} \cup \{v_{\ell'_1}, \dots, v_{\ell'_k}\}$. From the \(n^{\mathsf{k}}\) possible combinations, there are \(\binom{\mathsf{k}}{k}\) distinct sets of the form \(\{v_{\ell_1}, \dots, v_{\ell_k}\}\). For any fixed set \(\{v_{\ell_1}, \dots, v_{\ell_k}\}\), there are \(\binom{k}{2k - \mathsf{k}}\) ways to determine the intersection \(\{v_{\ell_1}, \dots, v_{\ell_k}\} \cap \{v_{\ell'_1}, \dots, v_{\ell'_k}\}\), which uniquely defines the second set \(\{v_{\ell'_1}, \dots, v_{\ell'_k}\}\). Therefore, there are \(\binom{\mathsf{k}}{k} \cdot \binom{k}{2k - \mathsf{k}}\) distinct ways to partition the union \(\{v_{\ell_1}, \dots, v_{\ell_k}\} \cup \{v_{\ell'_1}, \dots, v_{\ell'_k}\}\) into the two sets. 

There are $k!$ tuples for each of the set $\{v_{\ell_1}, \dots, v_{\ell_k}\}$ and $k!$ tuples for each of the set $\{v_{\ell'_1}, \dots, v_{\ell'_k}\}$. Hence,  
the number of such pairs \(\mathcal{W}, \mathcal{W'}\) such that \(|\{v_{\ell_1}, \dots, v_{\ell_k}\} \cap \{v_{\ell'_1}, \dots, v_{\ell'_k}\}| \geq 2\) is no more than:
$$\sum_{\mathsf{k} = k}^{2k - 2} n^{\mathsf{k}} \cdot \binom{\mathsf{k}}{k} \cdot \binom{k}{2k - \mathsf{k}} \cdot (k!)^2 = O(n^{2k - 2}).$$
\end{proof}

We are now ready to give the $\ell_2$-error of our estimator.

\begin{theorem}
The $\ell_2$-error of the estimator $\tilde{\sG}_k(G)$ is $O(n^{k-1})$.
\end{theorem}
\begin{proof}
The squared \(\ell_2\)-error is the sum of the variance and the squared bias. From Lemma \ref{lem:bias}, we have established that the bias is zero. Therefore, the remainder of this proof will focus on analyzing the variance of our estimator.

We first examine the variance of \(\tilde{W}(\mathcal{W}, \sG)\). This random variable is a product of a linear function involving \(O(k^2) = O(1)\) Bernoulli variables. Consequently, we can deduce that the variance of \(\tilde{W}(\mathcal{W}, \sG)\) is bounded by some constant \(C\). Applying the Cauchy-Schwarz inequality, we obtain that \(\mathrm{Cov}(\tilde{W}(\mathcal{W}, \sG), \tilde{W}(\mathcal{W}', \sG)) \leq C\).

From Lemma \ref{lem23}, we obtain that:
\begin{eqnarray*}
& & \mathrm{Var}[\tilde{\sG}_k(G)] \\ & = & \frac{1}{(A(\sG))^2} [\sum_{\mathcal{W} \in \mathcal{D}} \mathrm{Var} [\tilde{W}(\mathcal{W}, \sG)] + \\
& & ~~~~~~~~~~~\sum_{\mathcal{W} \in \mathcal{D}} \mathrm{Cov} [\tilde{W}(\mathcal{W}, \sG), \tilde{W}(\mathcal{W}', \sG)] ] \\
& \leq & \frac{1}{(A(\sG))^2} \left[ C \cdot n^k + C \cdot O(n^{2k - 2}) \right] = O(n^{2k - 2}).
\end{eqnarray*}
\end{proof}

\section{LOWER BOUND FOR NON- INTERACTIVE ALGORITHMS}

In this section, we aim to demonstrate that there exists a class of graphs for which any non-interactive local differentially private algorithm that attempts to count cliques of size \(k\) incurs an \(\ell_2\)-error of \(\Omega(n^{k - 1})\). Due to the technical complexity of the proof and space constraints, we outline the key proof ideas here and provide the full proof in the Appendix.

In this section, we assume, without loss of generality, that \( n \) is an integer divisible by three. The class of graphs under consideration is defined as follows.

For each $\mathbf{X} \in \{0,1\}^{n/3 \times n/3}$, $\mathbf{\mu} = \{\mu_1, \dots, \mu_{n/3}\} \in \{0,1\}^{n/3}$, and $\mathbf{\upsilon} = \{\upsilon_1, \dots, \upsilon_{n/3}\} \in \{0,1\}^{n/3}$, we construct the graph $\mathsf{G}_k^{\mathbf{\mu, \upsilon}}(\mathbf{X})$ by the following steps:
\begin{enumerate}
    \item We have $k$ set of nodes $\mathsf{U}, \mathsf{Y}, \mathsf{W}_1, \dots, \mathsf{W}_{k - 2}$. All of the sets have size $n / 3$. Let $\mathsf{U} = \{\mathsf{u}_1, \dots, \mathsf{u}_{n/3}\}$, $\mathsf{Y} = \{\mathsf{y}_1, \dots, \mathsf{y}_{n/3}\}$, and $\mathsf{W}_p = \{\mathsf{w}_{p,1}, \dots, \mathsf{w}_{p, n/3}\}$ for all $1 \leq p \leq k - 2$. The number of nodes in the graph is then equal to $k \cdot n / 3$.
    \item We have $\{\mathsf{u}_i, \mathsf{y}_j\} \in E$ if $\mathbf{X}_{i,j} = 1$.
    \item We have $\{\mathsf{u}_i, \mathsf{w}_{p,j}\} \in E$ for all $p,j$ if $\mathbf{\mu}_i = 1$.
    \item We have $\{\mathsf{w}_{p,i}, \mathsf{w}_{q,j}\} \in E$ for all $i$, $j$, and $p \neq q$.
    \item We have $\{\mathsf{y}_i, \mathsf{w}_{p,j}\} \in E$ for all $p,j$ if $\upsilon_i = 1$.
\end{enumerate}

Our gadget is shown as in Figure \ref{fig2}.
\begin{figure}[h]
    \centering
\includegraphics[width=0.7\linewidth]{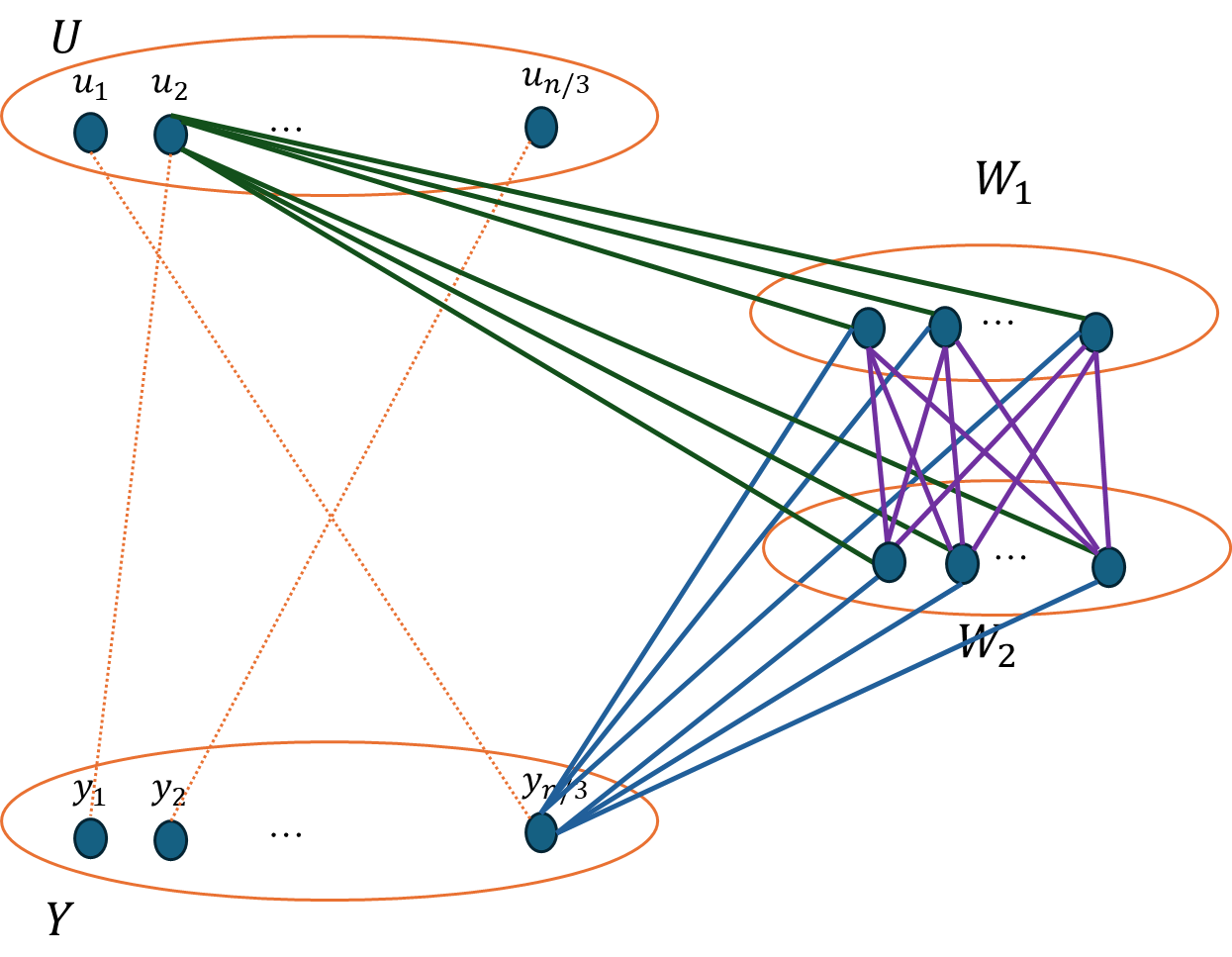}
    \caption{The gadget we use to show the lower bound of $\ell_2$-error when counting $K_4$.}
    \label{fig2}
\end{figure}

The main result of this section is as follows. The proof of this theorem is provided in the Appendix.
\begin{theorem} Let $\mathsf{n} = \frac{k \cdot n}{3}$ be the number of nodes in the graph $\mathsf{G}_k ^{\mathbf{\mu, \upsilon}}(\mathbf{X})$.
    There exists no non-interactive $\epsilon$-differentially private algorithm which can estimate the $K_k$ count in the graph $\mathsf{G}_k ^{\mathbf{\mu, \upsilon}}(\mathbf{X})$ with expected $\ell_2$-error in $o(\mathsf{n}^{k - 1})$.
\end{theorem}
Informally, it is shown in \cite{eden2023triangle} that for any differentially private algorithm \(\mathcal{A}\), there exists \(\mathsf{N} = \Theta(n^2)\), and a set of parameters \(\mathbf{X}, \mathbf{\mu}_1, \dots, \mathbf{\mu}_\mathsf{N}, \mathbf{\upsilon}_1, \dots, \mathbf{\upsilon}_\mathsf{N}\), such that the error in estimating the number of triangles for \textit{many} graphs in the set \(\{\mathsf{G}_3 ^{\mathbf{\mu}_1, \mathbf{\upsilon}_1}(\mathbf{X}), \dots, \mathsf{G}_3 ^{\mathbf{\mu}_\mathsf{N}, \mathbf{\upsilon}_\mathsf{N}}(\mathbf{X})\}\) is \(\Omega(n^2)\). 

We show that the number of \(K_k\) cliques in \(\mathsf{G}_k^{\mathbf{\mu, \upsilon}}(\mathbf{X})\) is \((n/3)^{k-3}\) times the number of triangles in \(\mathsf{G}_3^{\mathbf{\mu, \upsilon}}(\mathbf{X})\). If there existed a local differentially private algorithm that could estimate the number of \(K_k\) cliques in \(\mathsf{G}_k^{\mathbf{\mu, \upsilon}}(\mathbf{X})\) with \(\ell_2\)-error of \(o(n^{k - 1})\), we could divide the result by \((n/3)^{k-3}\) to obtain an estimator for the number of triangles in \(\mathsf{G}_3^{\mathbf{\mu, \upsilon}}(\mathbf{X})\) with an \(\ell_2\)-error of \(o(n^2)\). This contradicts the result in \cite{eden2023triangle}.

\section{LOWER BOUND FOR ANY ALGORITHM}

In this section, we explore the lower bound of estimating the number of graphlets with size $k$ under any local differentially private algorithm. Specifically, we give the lower bounds for the estimation of the number of cycles with length \(k\) in a graph \(G\), denoted by \(C_k(G)\).

Suppose the number of nodes $n$ is even. Let \(\mathbf{x} = (x_1, \dots, x_{n/2})\) be a bit vector of length \(n/2\). We define a graph \(G^{\mathbf{x}} = (V, E^{\mathbf{x}})\) as follows:
\begin{enumerate}
    \item Set \(V = \{v_1, \dots, v_n\}\).
    \item For \(u, v \in V\) where \(u \neq v\) and \(\{u, v\} \neq \{v_{2i - 1}, v_{2i}\}\) for all \(1 \leq i \leq n/2\), we have \(\{u, v\} \in E^\mathbf{x}\). 
    \item For $1 \leq i \leq n/2$, we include \(\{v_{2i-1}, v_{2i}\} \in E^\mathbf{x}\) if \(x_i = 1\). Let \(E_0^{\mathbf{x}} = \{\{v_{2i - 1}, v_{2i}\}: 1 \leq i \leq n/2\} \cap E^\mathbf{x}\). That is $|E_0^\mathbf{x}| = |\mathbf{x}|$.
\end{enumerate}
To establish a lower bound for the expected error, let us assume that all edges in \(G^{\mathbf{x}}\), except those in \(E_0^{\mathbf{x}}\), are non-sensitive and publicly known. It is understood that the expected error is lower when some edges are publicly known compared to when all edges are sensitive \cite{eden2023triangle}. Hence, this assumption can be utilized to derive a lower bound for the expected error. By the assumption, the set of possible graphs is then \(\mathcal{G} := \{G^{\mathbf{x}} : \mathbf{x} \in \{0, 1\}^{n/2}\}\).

For a subset \(E_p \subseteq E_0^{\mathbf{x}}\) of size \(p\), let \(\mathsf{C}_p\) denote the number of $k$-cycles in $G^{\mathbf{x}}$ that include all edges in \(E_p\). Since all edges in \(E_0^{\mathbf{x}}\) are identical, the number of $k$-cycles that use \(p\) edges from \(E_0^{\mathbf{x}}\) is \(\binom{|\mathbf{x}|}{p} \cdot \mathsf{C}_p\). Therefore, the total number of $k$-cycles in \(G^{\mathbf{x}}\) is given by:
\begin{equation}
    f(|\mathbf{x}|) = C_k(G^{\mathbf{0}}) + \sum_{p = 1}^{|\mathbf{x}|} \binom{|\mathbf{x}|}{p} \cdot \mathsf{C}_p. \label{eqn2}
\end{equation} 
\begin{lemma}
    For any \(\mathbf{x}, \mathbf{x'}\) and $D \geq 0$ such that \(|\mathbf{x}| - |\mathbf{x'}|  = D\), we have 
    \[
    C_k\left(G^{\mathbf{x}}\right) - C_k\left(G^{\mathbf{x'}}\right) = \Omega\left(n^{k - 2} \cdot D\right).
    \]
    \label{lemBound}
\end{lemma}
\begin{proof}
     We have:
     \begin{eqnarray*}
         & & C_k\left(G^{\mathbf{x}}\right) - C_k\left(G^{\mathbf{x'}}\right) \\ &=& \left( C_k(G^{\mathbf{0}}) + \sum_{p = 1}^{|\mathbf{x}|} \binom{|\mathbf{x}|}{p} \cdot \mathsf{C}_p \right) \\ & & ~~~~- \left( C_k(G^{\mathbf{0}}) + \sum_{p = 1}^{|\mathbf{x'}|} \binom{|\mathbf{x'}|}{p} \cdot \mathsf{C}_p \right) \\
         &=& (|\mathbf{x}| - |\mathbf{x'}|) \cdot \mathsf{C}_1 \\
         & & ~~~~ + \left( \sum_{p = 2}^{|\mathbf{x}|} \binom{|\mathbf{x}|}{p} \cdot \mathsf{C}_p - \sum_{p = 2}^{|\mathbf{x'}|} \binom{|\mathbf{x'}|}{p} \cdot \mathsf{C}_p \right) \\
         &\geq& (|\mathbf{x}| - |\mathbf{x'}|) \cdot \mathsf{C}_1.
     \end{eqnarray*}
     Because the number of cycles that use one edge in \(E_0^{\mathbf{x}}\) is \(\Theta(n^{k-2})\), we obtain the lemma statement.
\end{proof}
\textcolor{black}{The primary theorem in this section relies on the following lower bound result from \cite{joseph2019role}.}
\begin{theorem} [Theorem 5.3 from version 2 of the ArXiv preprint \cite{joseph2019role}] \textcolor{black}{Let the sensitive information of users \( v_1, \dots, v_n \) be represented by \( x_1, \dots, x_n \in \{0,1\} \). No locally differentially private algorithm can compute \( \sum_{i=1}^{n} x_i \) with an \(\ell_2\)-error of \( o(\sqrt{n}) \).} \label{thm:Joseph} 
\end{theorem}

Now, we are ready to present the main theorem of this section. 

\begin{theorem}
    There is no local \(\epsilon\)-edge differential privacy algorithm that can estimate the number of \(C_k\) with an expected \(\ell_2\)-loss of \(o(n^{k - 1.5})\). 
\end{theorem}

\begin{proof}
Assume, for contradiction, that such an algorithm exists. We call this algorithm Algorithm \(\mathcal{A}\). Suppose we give a graph \(G^{\mathbf{x}} \in \mathcal{G}\) to Algorithm \(\mathcal{A}\) to count the number of \(C_k\). Next, we introduce an algorithm \(\mathcal{B}\) designed to estimate \(|\mathbf{x}|\). Recall the function $f$ defined in (\ref{eqn2}). Algorithm \(\mathcal{B}\) outputs \(q\) such that \(f(q)\) is closest to the output of Algorithm \(\mathcal{A}\) on $G^{\mathbf{x}}$.

If the expected \(\ell_2\)-loss of Algorithm \(\mathcal{A}\), which is a local \(\epsilon\)-edge differential privacy algorithm, is \(o(n^{k - 1.5})\), then, by Lemma \ref{lemBound}, Algorithm \(\mathcal{B}\), also a local \(\epsilon\)-edge differential privacy algorithm, must have an estimation error of \(o(\sqrt{n})\). \textcolor{black}{However, this contradicts with Theorem \ref{thm:Joseph}.}
\end{proof}

\section{EXPERIMENTAL RESULTS}

\subsection{Settings}
In this section, we discuss our experimental results, which are used to confirm the validity of our theoretical results.    Our experiment settings are as follows: 
\paragraph{Evaluation Method} 
\textcolor{black}{For each set of parameters, we generate a graph \( G \) where the actual number of subgraphs is denoted as \( \mathsf{S}(G) \). We then execute our algorithm ten times, obtaining estimates \( i_1, \dots, i_{10} \). Then we evaluate the precision using the root mean square error, which can be computed as:
\[
\sqrt{\sum_{t=1}^{10} (i_t - \mathsf{S}(G))^2}.
\]
We also evaluate our algorithm using the relative root mean square error, which can be computed as:
\[
\frac{\sqrt{\sum_{t=1}^{10} (i_t - \mathsf{S}(G))^2}}{\mathsf{S}(G)}.
\]}

\paragraph{Input graphs} Recall that $n$ is the size of the input graph and $k$ is the size of graphlets. The computation complexity of our algorithm is $O(n^k)$. Hence, we cannot conduct experiments on a large input graphs at this state. We therefore choose to conduct experiments on synthetic graphs with size 10, 20, \dots, 100 in this paper.

The input graphs are derived from the Barabási–Albert model \cite{albert2002statistical} and the stochastic block model \cite{holland1983stochastic}, selected to represent two distinct types of social networks. The stochastic block model captures networks with multiple clusters, while the Barabási–Albert model illustrates networks characterized by a small number of central nodes. 

Given the relatively small number of nodes, we configured the stochastic block model with two blocks, each containing an equal number of nodes, $n/2$. The probability of a link between two nodes within the same block is set to 0.25, while the probability of a link between nodes in different blocks is 0.05. In the Barabási–Albert model, the degree of each newly added node is set to $n / 5$.

\paragraph{Graphlet} In this experiment, we focus on counting four-node cycles. Based on our algorithm and the corresponding theoretical results, we expect that the experimental outcomes would not significantly vary for different types of graphlets. Therefore, we selected the smallest graphlet to minimize the computation time of our algorithm.

\paragraph{Benchmark Algorithm} Apart from the work by Imola et al. \cite{imola2022differentially}, which reports the number of four-length cycles under an assumption weaker than local differential privacy, and the work by He et al. \cite{he2024butterfly}, which assumes that the input graph is bipartite, we are not aware of any algorithm specifically designed for counting four-length cycles. Therefore, in this experiment, we choose to compare our algorithm with the classical randomized response technique.

\paragraph{Privacy Budget} We set the privacy budget to 1 and 5 in our experiment, which may seem relatively high. However, due to the small number of nodes in this experiment, a higher privacy budget is necessary to ensure meaningful results. We believe that for larger graphs, a significantly lower privacy budget would suffice without compromising the quality of the publication.

The code for our experiments are available at \url{https://github.com/gcwsilver/Counting-Graphlets-of-Size-k-under-Local-Differential-Privacy}.

\subsection{Results}

\begin{figure}[htbp]
    \centering
    \begin{subfigure}{0.23\textwidth}
        \centering
        \includegraphics[width=\textwidth]{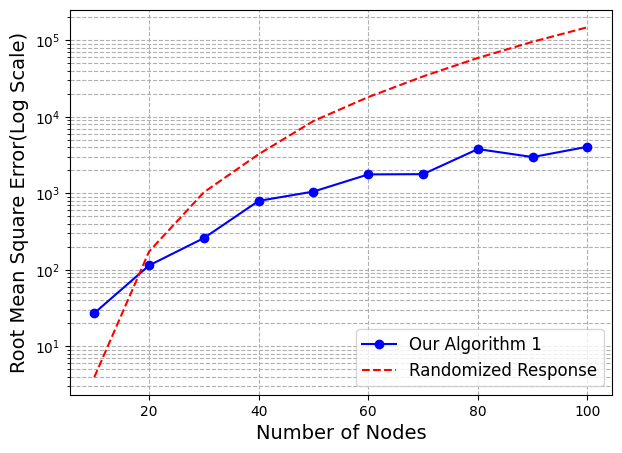}
        \caption{}
        \label{fig:sub1}
    \end{subfigure}
    \hfill
    \begin{subfigure}{0.23\textwidth}
        \centering
        \includegraphics[width=\textwidth]{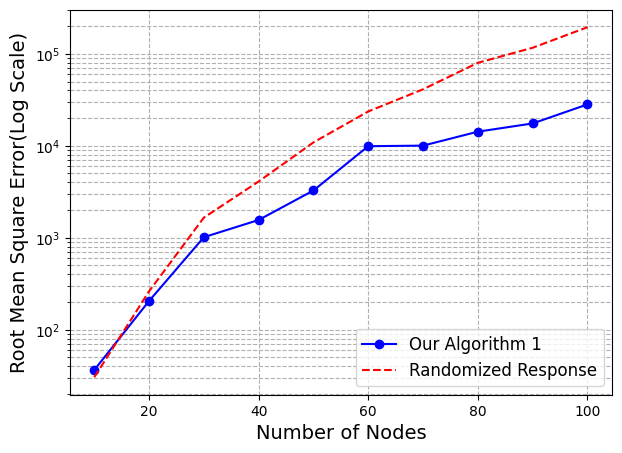}
        \caption{}
        \label{fig:sub3}
    \end{subfigure}
        \begin{subfigure}{0.23\textwidth}
        \centering
        \includegraphics[width=\textwidth]{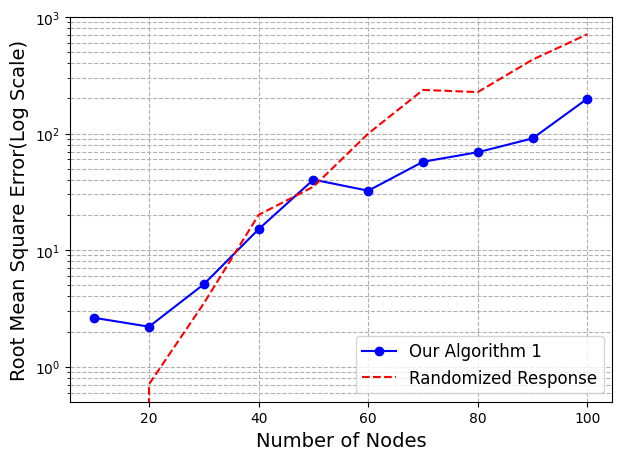}
        \caption{}
        \label{fig:sub2}
    \end{subfigure}
    \hfill
    \begin{subfigure}{0.23\textwidth}
        \centering
        \includegraphics[width=\textwidth]{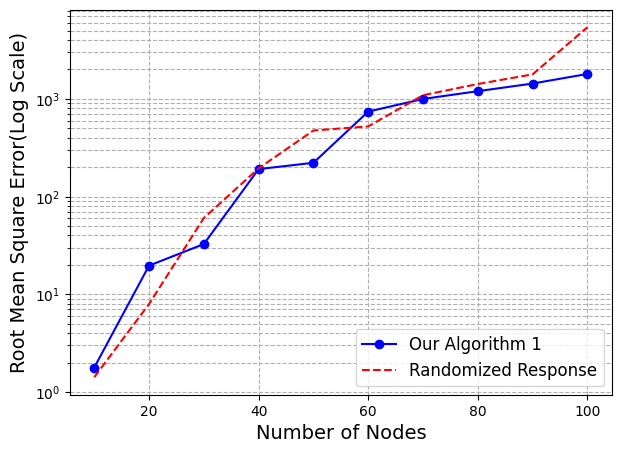}
        \caption{}
        \label{fig:sub4}
    \end{subfigure}
    
    \caption{Root mean square error of our Algorithm~\ref{alg1} compared with the  randomized response mechanism (a) in the stochastic block model when the privacy budget $\epsilon = 1$, (b) in the Barabási–Albert model when $\epsilon = 1$, (c) in the stochastic block model when $\epsilon = 5$, and (d) in the Barabási–Albert model when $\epsilon = 5$. We omit the result for the randomized response in (c) when \( n = 10 \) as the error is zero, making it impossible to display on a log scale plot. }
    \label{fig3}
\end{figure}

We give the root mean square error of our Algorithm~\ref{alg1} compared with the classical randomized response in Figure~\ref{fig3}. We found that our algorithm significantly better than the randomized rounding in almost all of the settings. The figures indicate that our algorithm shows greater improvement in the stochastic block model, demonstrating its clear advantage for clustered social networks. Additionally, the results reveal that the improvement increases as the number of nodes grows and the privacy budget \( \epsilon \) decreases. This suggests that in practical scenarios with a much larger number of nodes and a smaller privacy budget, our algorithm's improvement will be even more significant than what is shown in the plots. Specifically, in the stochastic block model, our algorithm achieves an improvement of approximately 36 times when the number of nodes is 100 and the privacy budget is one.

The results, showing that our improvements are greater for a larger number of nodes, align with our theoretical findings. Specifically, the expected \( \ell_2 \)-error of our algorithm is \( O(n^{k - 1}) \), while the expected \( \ell_2 \)-error for the randomized response is \( \Theta(n^{k}) \).

\begin{figure}[htbp]
    \centering
    \begin{subfigure}{0.23\textwidth}
        \centering
        \includegraphics[width=\textwidth]{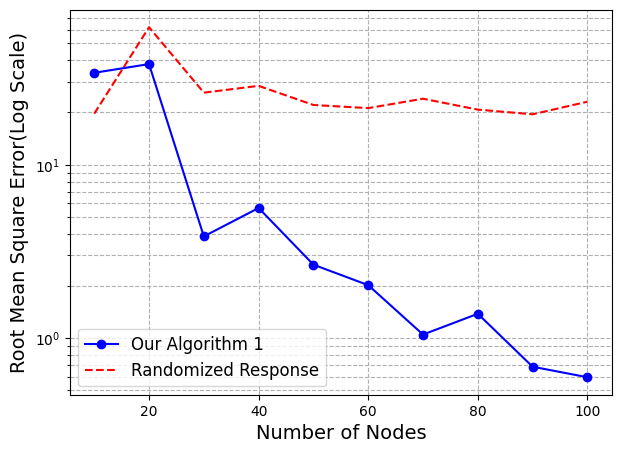}
        \caption{}
        \label{fig:sub5}
    \end{subfigure}
    \hfill
    \begin{subfigure}{0.23\textwidth}
        \centering
        \includegraphics[width=\textwidth]{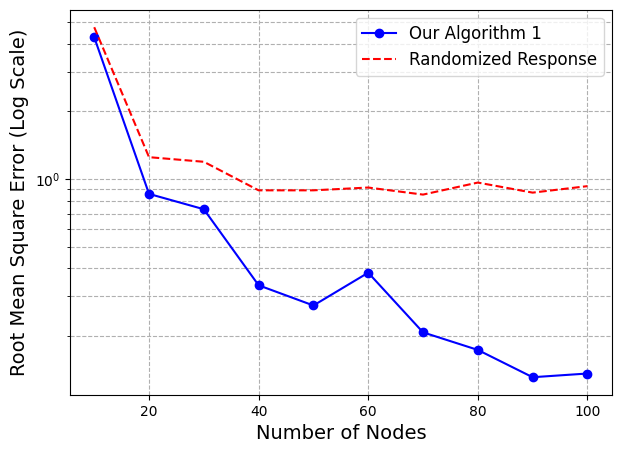}
        \caption{}
        \label{fig:sub7}
    \end{subfigure} \\
    \begin{subfigure}{0.23\textwidth}
        \centering
        \includegraphics[width=\textwidth]{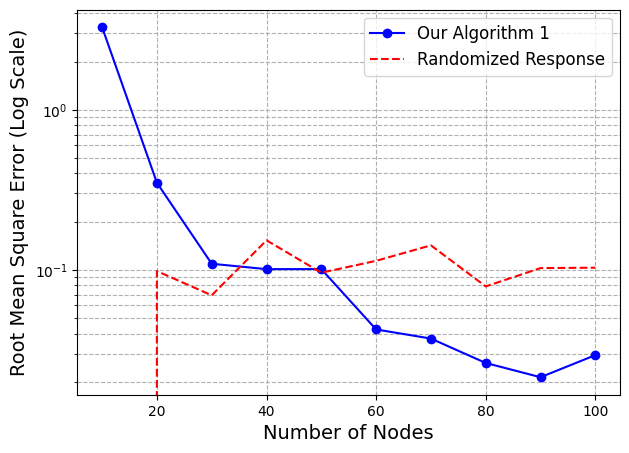}
        \caption{}
        \label{fig:sub6}
    \end{subfigure}
    \hfill
    \begin{subfigure}{0.23\textwidth}
        \centering
        \includegraphics[width=\textwidth]{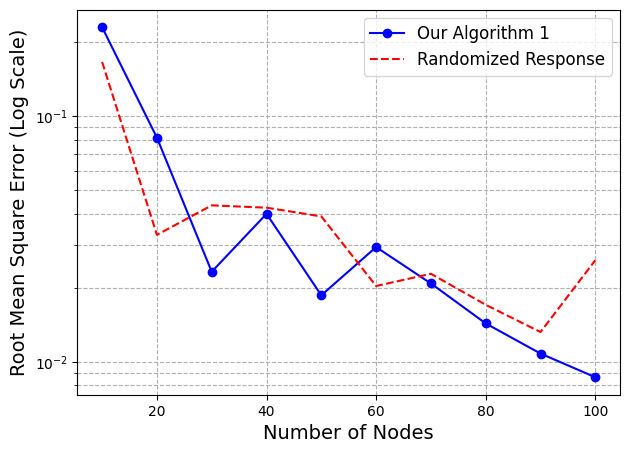}
        \caption{}
        \label{fig:sub8}
    \end{subfigure}

    \caption{Relative root mean square error of our Algorithm~\ref{alg1} compared with the  randomized response mechanism in the stochastic block model (a) in the stochastic block model when the privacy budget $\epsilon = 1$, (b) in the Barabási–Albert model when $\epsilon = 1$, (c) in the stochastic block model when $\epsilon = 5$,  and (d) in the Barabási–Albert model when $\epsilon = 5$. We omit the result for the randomized response in (c) when \( n = 10 \) as the error is zero, making it impossible to display on a log scale plot. }
    \label{fig4}
\end{figure}

In Figure~\ref{fig4}, we present the relative root mean square error, calculated as the relative mean square error divided by the average value. The plots clearly demonstrate that the relative error decreases as the number of nodes increases. Even with as few as 100 nodes, the plot confirms that our algorithm produces meaningful results. Specifically, the relative error is below 0.6 when the privacy budget $\epsilon = 1$, and it drops to less than 0.03 when $\epsilon = 5$. This decline in relative error with increasing nodes aligns with our theoretical predictions. Since the probability that any tuple of length four forms a cycle in the stochastic block model is $\Theta(1)$, the expected number of 4-cycles in graphs generated by the model increases as $\Theta(n^4)$. Meanwhile, the $\ell_2$-error of our algorithm scales as $\Theta(n^3)$.

Conversely, the plot shows that the relative error from the randomized response mechanism remains constant as the number of nodes increases. This outcome is also theoretically anticipated, as the $\ell_2$-error of the classical mechanism is $\Theta(n^4)$, matching the order of the expected number of 4-cycles.

\begin{figure}[htbp]
    \centering
    \begin{subfigure}{0.23\textwidth}
        \centering
        \includegraphics[width=\textwidth]{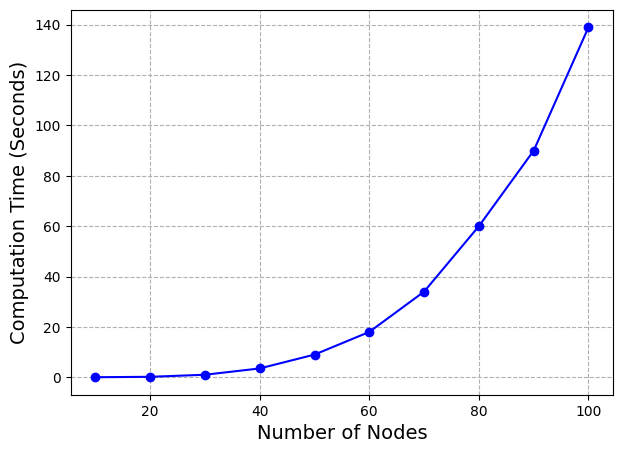}
        \caption{}
        \label{fig:sub9}
    \end{subfigure}
    \hfill
    \begin{subfigure}{0.23\textwidth}
        \centering
        \includegraphics[width=\textwidth]{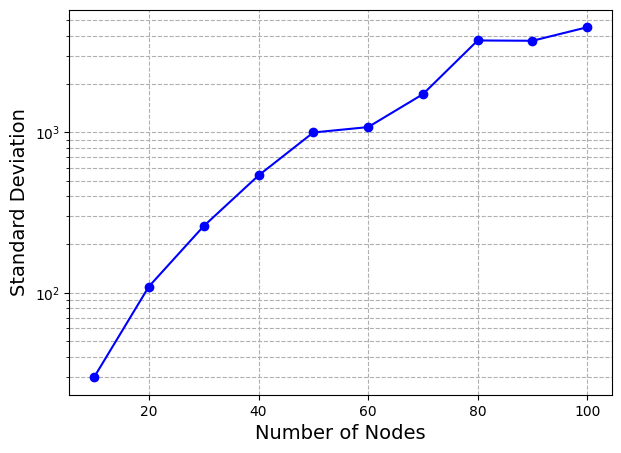}
        \caption{}
        \label{fig:sub10}
    \end{subfigure} 
        \caption{(a) Computation time of our Algorithm 1 as a function of the number of nodes (b) Standard deviation of our results for graphs derived from the stochastic block model when the privacy budget is set to one.}
    \label{fig:computation_time}
\end{figure}

Next, we address the computation time of our algorithm, which represents its main drawback. In Figure~\ref{fig:sub9}, we display the computation time for Algorithm~\ref{alg1} executed on Google Colab. The figure shows that the computation time increases rapidly, following an $O(n^4)$ growth pattern. For instance, the algorithm takes 139 seconds to compute the number of 4-cycles in a graph with 100 nodes. This suggests that the computation time may be too lengthy for practical applications.

On the other hands, we can optimize the counting of some specific subgraph based on the randomized response mechanism. As the results of the mechanism is a network, we can utilize subgraph counting algorithms (such as \cite{ahmed2015efficient,rahman2014graft}) to speed up the count. The results based on the mechanism is hence can be calculated in a much more efficient way.

In Figure \ref{fig:sub10}, we present the standard deviation for graphs generated under the stochastic block model when the privacy budget is set to one. The standard deviation closely matches the $\ell_2$-error depicted in Figure \ref{fig:sub1}, confirming that the error in our algorithm is solely due to variance, with no indication of bias.

\section{CONCLUSION AND FUTURE WORKS}

In this paper, we establish the lower bound for the $\ell_2$-error in the task of counting subgraphs of size $k$ under local differential privacy. We show that for a certain class of input graphs of size $n$, any locally differentially private algorithm must incur an $\ell_2$-error of at least $\Omega(n^{k - 1.5})$, and for any non-interactive locally differentially private algorithm, the $\ell_2$-error is at least $\Omega(n^{k - 1})$. Moreover, we match this lower bound by proposing a non-interactive, locally differentially private algorithm that achieves an $\ell_2$-error of $O(n^{k - 1})$ for counting any subgraph of size $k$. Our algorithm leverages the randomized response mechanism and employs a similar unbiased technique as the one used for triangle counting in \cite{eden2023triangle}. We generalize this approach by using subgraph automorphisms to extend the triangle counting method to any subgraphs.

The primary limitation of our algorithm is its computational complexity. Currently, the algorithm has a computation time of $O(n^k)$, which is impractically high for real-world applications. Nonetheless, we believe it provides a foundation for scalable and locally private counting. \textcolor{black}{We believe that incorporating the sampling technique proposed for non-private subgraph counting  \cite{ribeiro2021survey} could enhance our algorithm's speed. We anticipate that, with these techniques, our algorithm will achieve computation times comparable to state-of-the-art non-private counting methods.} 

\textcolor{black}{For the theoretical results in this work, we derive upper and lower bounds for the \(\ell_2\)-error in terms of the number of users \( n \). Here, we treat \(\epsilon\) as a constant, since in practical scenarios, \( 1/\epsilon \) is typically much smaller than \( n \) and rarely exceeds 100. However, expressing both upper and lower bounds in terms of \( n \) and \( \epsilon \) would offer deeper insights for future research. Therefore, we plan to derive such bounds in future work.}

\section*{Acknowledgments}

\textcolor{black}{The authors sincerely thank the anonymous reviewers for their valuable comments, which have greatly contributed to improving this paper. Vorapong Suppakitpaisarn receives partial support from KAKENHI Grants 21H05845 and 23H04377, as well as JST NEXUS Grant Y2024L0906031. Quentin Hillebrand is partially supported by KAKENHI Grant 20H05965, and by JST SPRING Grant Number JPMJSP2108. This research was partially conducted during Donlapark Ponnoprat's stay at The University of Tokyo, hosted by Masaaki Imaizumi. Nicha Hirankarn joined this project upon the introduction of Sira Srisawat (Chulalongkorn University). The authors would also like to take this opportunity to thank Masaaki Imaizumi and Sira Srisawat.}

\bibliographystyle{apalike}
\bibliography{references}

\section*{Checklist}

 \begin{enumerate}

 \item For all models and algorithms presented, check if you include:
 \begin{enumerate}
   \item A clear description of the mathematical setting, assumptions, algorithm, and/or model. [Yes, we outline the setting and the algorithm in Sections 2 and 3.]
   \item An analysis of the properties and complexity (time, space, sample size) of any algorithm. [Yes, we analyze the privacy of our algorithm in Section 3. However, we do not provide an explicit analysis of the time and space complexity, as they are straightforward.]
   \item (Optional) Anonymized source code, with specification of all dependencies, including external libraries. [Not Applicable]
 \end{enumerate}

 \item For any theoretical claim, check if you include:
 \begin{enumerate}
   \item Statements of the full set of assumptions of all theoretical results. [Not applicable. Our algorithm and lower bounds apply to the general case, so no specific assumptions are required.]
   \item Complete proofs of all theoretical results. [Yes. We provide the proofs for all theoretical results in Sections 3-5, and the supplementary material.]
   \item Clear explanations of any assumptions. [Not applicable. Our algorithm and lower bounds apply to the general case, so no specific assumptions are required.]     
 \end{enumerate}

 \item For all figures and tables that present empirical results, check if you include:
 \begin{enumerate}
   \item The code, data, and instructions needed to reproduce the main experimental results (either in the supplemental material or as a URL). [Yes. We provide the code and the instructions as a supplemental material. The data is synthesized from the code.]
   \item All the training details (e.g., data splits, hyperparameters, how they were chosen). [Not Applicable. We do not train any machine learning model in this work.]
         \item A clear definition of the specific measure or statistics and error bars (e.g., with respect to the random seed after running experiments multiple times). [Yes. We describe that in Section 6.]
         \item A description of the computing infrastructure used. (e.g., type of GPUs, internal cluster, or cloud provider). [Yes. We describe that in Section 6.]
 \end{enumerate}

 \item If you are using existing assets (e.g., code, data, models) or curating/releasing new assets, check if you include:
 \begin{enumerate}
   \item Citations of the creator If your work uses existing assets. [Not Applicable. We do not use any existing asset in this work.]
   \item The license information of the assets, if applicable. [Not Applicable. We do not use any existing asset in this work.]
   \item New assets either in the supplemental material or as a URL, if applicable. [Yes. We include the code in the supplemental material.]
   \item Information about consent from data providers/curators. [Not Applicable. We do not use any existing asset in this work.]
   \item Discussion of sensible content if applicable, e.g., personally identifiable information or offensive content. [Not Applicable. We do not collect any information in this work.]
 \end{enumerate}

 \item If you used crowdsourcing or conducted research with human subjects, check if you include:
 \begin{enumerate}
   \item The full text of instructions given to participants and screenshots. [Not Applicable. We do not conduct any experiment with human subjects.]
   \item Descriptions of potential participant risks, with links to Institutional Review Board (IRB) approvals if applicable. [Not Applicable. We do not conduct any experiment with human subjects.]
   \item The estimated hourly wage paid to participants and the total amount spent on participant compensation. [Not Applicable. We do not conduct any experiment with human subjects.]
 \end{enumerate}

 \end{enumerate}

\onecolumn
\aistatstitle{
Supplemental Materials for the Manuscript ``Counting Graphlets of Size $k$ under Local Differential Privacy''}

\section{MISSING PROOFS FOR SECTION 4}

We give details for our lower bound proof in Section 4. In the section, we discuss that there is a class of input graphs such that any non-interactive locally differentially private algorithm which counts the number of cliques with size $k$ must have an $\ell_2$ error of $\Omega(n^{k - 1})$. Let the input graph be $G$, and let the number of cliques in the graph be $K_k(G)$. 

\subsection{Result of \cite{eden2023triangle}}

Our proof will rely on the lower bound of error in counting triangles as established in \cite{eden2023triangle}. In the paper, the authors demonstrate that no non-interactive $\epsilon$-edge local differential privacy mechanism can count triangles with an expected $\ell_2$-error less than \(o(n^2)\). However, we cannot directly utilize this result as a black box for our proof. Instead, we derive our lower bound by leveraging the gadget described in the paper.

\subsubsection*{Gadget $G^{\mathbf{\mu}, \mathbf{\upsilon}}(\mathbf{X})$}
We begin by describing the gadget used in this paper. Without loss of generality, we assume that $3 \vert n$. Let \(\mathbf{\mu}, \mathbf{\upsilon} \in \{0,1\}^{n/3}\). Also, let \(\mathbf{X}\) be a \(0\)-\(1\) matrix of size \(\frac{n}{3} \times \frac{n}{3}\). We define the gadget, denoted by \(G^{\mathbf{\mu}, \mathbf{\upsilon}}(\mathbf{X})\), as follows:

\begin{enumerate}
    \item The graph is a tripartite graph. Let the sets of nodes in each partition be \(U = \{u_1, \dots, u_{n/3}\}\), \(Y = \{y_1, \dots, y_{n/3}\}\), and \(W = \{w_1, \dots, w_{n/3}\}\).
    \item For each pair $i,j$, there is an edge between \(u_i\) and \(y_j\) if \(\mathbf{X}_{i,j} = 1\); otherwise, there is no edge between them.
    \item For each $i$, there is an edge between \(u_i\) and \(w_j\) \emph{for all} $j$ if \(\mathbf{\mu}_i = 1\); otherwise, there is no edge between them.
    \item For each $i$, there is an edge between \(y_i\) and \(w_j\) \emph{for all} $j$ if \(\mathbf{\upsilon}_i = 1\); otherwise, there is no edge between them.
\end{enumerate}

\begin{figure}[h]
    \centering
\includegraphics[width=0.4\linewidth]{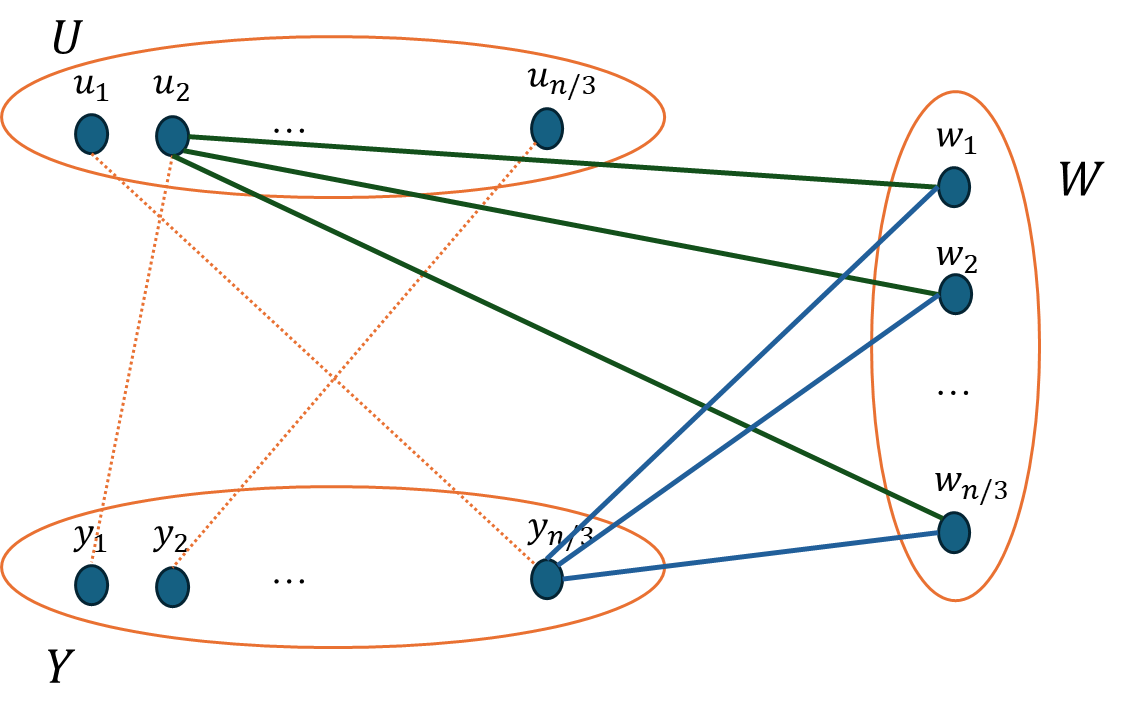}
    \caption{Gadget used in \cite{eden2023triangle} for showing the lower bound in $\ell_2$-error of estimating the number of triangles.}
    \label{fig1}
\end{figure}

We illustrate the gadget in Figure \ref{fig1}.

\subsubsection*{Edge Differentially Private Algorithm $\mathcal{C}$}

For each node \( v \), let \( \mathbf{a}_v \) represent the adjacency vector of \( v \) in the graph $G^{\mathbf{\mu}, \mathbf{\upsilon}}(\mathbf{X})$. A non-interactive \(\epsilon\)-edge differential privacy algorithm, denoted by \(\mathcal{C}\), can be described as follows:

\begin{enumerate}
    \item Each node \( v \) applies an \(\epsilon\)-differential privacy algorithm, denoted by \(\mathcal{L}_v\) (referred to as a randomizer), to its adjacency vector \( \mathbf{a}_v \). The node then sends the result \(\mathcal{L}_v(\mathbf{a}_v)\) to a central server.
    \item The central server applies an aggregator function \(\mathcal{A}\) to the collected results from all randomizers and outputs \(\mathcal{A}\left( \langle \mathcal{L}_v(\mathbf{a}_v) \rangle_{v \in V} \right)\) as the final result.
\end{enumerate}
Let us now recall the graph \( G^{\mathbf{\mu,\upsilon}}(\mathbf{X}) \). For a node \( u_i \), we denote its adjacency vector in the graph as \( \mathbf{a}_{u_i}(\mathbf{\mu, \upsilon, X}) \). Given any vectors \( \mathbf{\mu}, \mathbf{\mu}', \mathbf{\upsilon}, \mathbf{\upsilon}' \) where \( \mathbf{\mu}_i = \mathbf{\mu}'_i \), we have \( \mathbf{a}_{u_i}(\mathbf{\mu, \upsilon, X}) = \mathbf{a}_{u_i}(\mathbf{\mu', \upsilon', X}) \). Therefore, we can simplify the notation for \( \mathbf{a}_{u_i}(\mathbf{\mu, \upsilon, X}) \) to \( \mathbf{a}_{u_i}(\mathbf{\mu}_i, \mathbf{X}) \). Similarly, \( \mathbf{a}_{v_i}(\mathbf{\mu, \upsilon, X}) \) can be abbreviated as \( \mathbf{a}_{v_i}(\mathbf{\upsilon}_i, \mathbf{X}) \), and, by the definition, \( \mathbf{a}_{w_i}(\mathbf{\mu, \upsilon, X}) \) can be abbreviated as \( \mathbf{a}_{w_i}(\mathbf{\mu}, \mathbf{\upsilon}) \) for all $w_i \in W$. Since the computation of $w_i$ do not involve any private information, we can assume that those computations are done at the central server.

\subsubsection*{Edge Differentially Private Algorithm $\mathcal{D}_\mathcal{C}$}

Let us assume that the algorithm $\mathcal{C}$ is designed to estimate the number of triangles in a given graph. Additionally, suppose that for any graph $G^{\mathbf{\mu, \upsilon}}(\mathbf{X})$, the edges between the node sets $U \cup Y$ and $W$ (described by $\mathbf{\mu, \upsilon}$) are not private, while the edges between the node sets $U$ and $Y$ (described by $\mathbf{X}$) are private. For a fixed private matrix $\mathbf{X}$, the authors of \cite{eden2023triangle} have designed a $2\epsilon$-differentially private algorithm, denoted as Algorithm $\mathcal{D}_\mathcal{C}$, which can estimate the number of triangles in $\mathsf{N} = \Theta(n^2)$ graphs in the set $\{G^{\mathbf{\mu, \upsilon}}(\mathbf{X}): \mu, \upsilon \in \{0,1\}^{n/3}\}$ based on the functions $\mathcal{L}_v$ and $\mathcal{A}$ within the algorithm $\mathcal{C}$. Let the set of these $\Theta(n^2)$ graphs be $\mathcal{S}_{\mathbf{X}}$. The algorithm $\mathcal{D}_\mathcal{C}$ can be described as follows:

\begin{enumerate}
    \item For each \( u_i \in U \), the node \( u_i \) applies the algorithm \( \mathcal{L}_{u_i} \) to the adjacency vectors \( \mathbf{a}_{u_i}(0, \mathbf{X}) \) and \( \mathbf{a}_{u_i}(1, \mathbf{X}) \). The node then sends the results to the central server. Let these results be \( \mathcal{L}_{u_i}(0) \) and \( \mathcal{L}_{u_i}(1) \). Since the node publishes two results, this process is \( 2\epsilon \)-differentially private.  
    \item For each \( y_i \in Y \), the node \( y_i \) applies the algorithm \( \mathcal{L}_{y_i} \) to the adjacency vectors \( \mathbf{a}_{y_i}(0, \mathbf{X}) \) and \( \mathbf{a}_{y_i}(1, \mathbf{X}) \). The node then sends the results to the central server. Let these results be \( \mathcal{L}_{y_i}(0) \) and \( \mathcal{L}_{y_i}(1) \). Since the node publishes two results, this process is \( 2\epsilon \)-differentially private.
    \item Then, for all graphs in \( \mathcal{S}_\mathbf{X} = \{ G^{\mathbf{\mu^{(1)}, \upsilon^{(1)}}}(\mathbf{X}), \dots, G^{\mathbf{\mu^{(\mathsf{N})}, \upsilon^{(\mathsf{N})}}}(\mathbf{X}) \} \):
    \begin{enumerate}
        \item Each user \( w_i \) first applies the randomized algorithm \( \mathcal{L}_{w_i} \) to its adjacency vector \( \mathbf{a}_{w_i}(\mathbf{\mu}^{(1)}, \mathbf{\upsilon}^{(1)}, \mathbf{X}), \dots, \mathbf{a}_{w_i}(\mathbf{\mu}^{(\mathsf{N})}, \mathbf{\upsilon}^{(\mathsf{N})}, \mathbf{X}) \). Then, the user sends the result to the central server. Note that all edges incident to \( w_i \) are unrelated to the private matrix \( \mathbf{X} \), ensuring that this process does not leak any privacy.
        \item The central server then applies the algorithm \( \mathcal{A} \) to  \( \langle \langle \mathcal{L}_{u_i}  (\mathbf{\mu}_i^{(t)}) \rangle_{i = 1}^{n/3} \), \( \langle \mathcal{L}_{y_i}(\mathbf{\upsilon}^{(t)}_i) \rangle_{i = 1}^{n/3}, \langle \mathbf{a}_{w_i}(\mu^{(t)}, \upsilon^{(t)}) \rangle_{i = 1}^{n/3} \rangle \) for each $t \in [\mathsf{N}]$. The result obtained from this step is denoted as \( \mathcal{D}_\mathcal{C}(\mu^{(1)}, \upsilon^{(1)}, \mathbf{X}), \dots, \mathcal{D}_\mathcal{C}(\mu^{(\mathsf{N})}, \upsilon^{(\mathsf{N})}, \mathbf{X}) \).
    \end{enumerate}
\end{enumerate}

The authors of \cite{eden2023triangle} show the following result in their paper.
\begin{theorem}[\cite{eden2023triangle}] Let $n$ be the number of nodes in the graph $G^{\mu^{(t)}, \nu^{(t)}} (\mathbf{X})$. \label{thm41}
    There exists $\mathsf{c} > 0$, $\mathsf{N} = \mathsf{c} \cdot n^2$,  such that no non-interactive  $\epsilon$-differentially private algorithms $\mathcal{C}$ where, for all $\mathbf{X}$, $\mu^{(1)}, \dots, \mu^{(\mathsf{N})}$, $\upsilon^{(1)}, \dots, \upsilon^{(\mathsf{N})}$,
    \begin{equation}
        \Pr[\left|\left\{ t \in \{1, \dots, \mathsf{N} \} : \left|\mathcal{D}_\mathcal{C}(\mu^{(t)}, \nu^{(t)}, \mathbf{X}) - K_3\left(G^{\mu^{(t)}, \nu^{(t)}} (\mathbf{X})\right)\right| \leq \frac{n^2}{12} \right\} \right| > \frac{\mathsf{N}}{5184}] > \frac{1}{6}. \label{eqn3}
    \end{equation}
\end{theorem}

\subsection{Our Construction}

We define our construction in this subsection.

\subsubsection*{Graph $\mathsf{G}_k^{\mathbf{\mu, \upsilon}}(\mathbf{X})$}
For each $G^{\mathbf{\mu, \upsilon}}(\mathbf{X})$, we construct the graph $\mathsf{G}_k^{\mathbf{\mu, \upsilon}}(\mathbf{X})$ by the following steps:
\begin{enumerate}
    \item We have $k$ set of nodes $\mathsf{U}, \mathsf{Y}, \mathsf{W}_1, \dots, \mathsf{W}_{k - 2}$. All of the sets have size $n / 3$. Let $\mathsf{U} = \{\mathsf{u}_1, \dots, \mathsf{u}_{n/3}\}$, $\mathsf{Y} = \{\mathsf{y}_1, \dots, \mathsf{y}_{n/3}\}$, and $\mathsf{W}_p = \{\mathsf{w}_{p,1}, \dots, \mathsf{w}_{p, n/3}\}$ for all $1 \leq p \leq k - 2$. The number of nodes in the graph, denoted by $\mathsf{n}$, is then equal to $k \cdot n / 3$.
    \item We have $\{\mathsf{u}_i, \mathsf{y}_j\} \in E$ if $\mathbf{X}_{i,j} = 1$.
    \item We have $\{\mathsf{u}_i, \mathsf{w}_{p,j}\} \in E$ for all $p,j$ if $\mathbf{\mu}_i = 1$.
    \item We have $\{\mathsf{w}_{p,i}, \mathsf{w}_{q,j}\} \in E$ for all $i$, $j$, and $p \neq q$.
    \item We have $\{\mathsf{y}_i, \mathsf{w}_{p,j}\} \in E$ for all $p,j$ if $\upsilon_i = 1$. 
\end{enumerate}

Our gadget is shown as in Figure \ref{fig2}.
\begin{figure}[h]
    \centering
\includegraphics[width=0.4\linewidth]{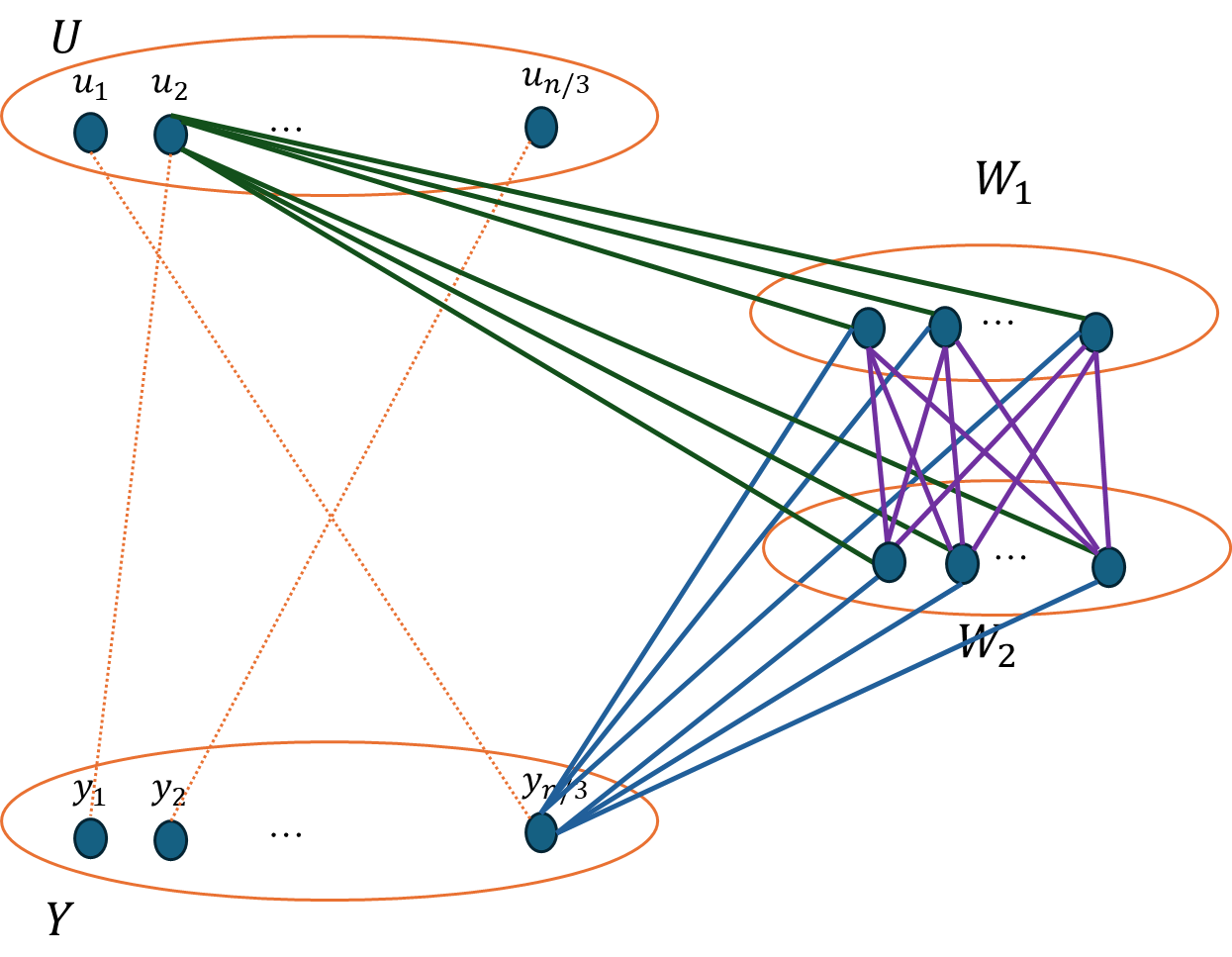}
    \caption{The gadget we use to show the lower bound of $\ell_2$-error when counting $K_4$.}
    \label{fig2}
\end{figure}

By the construction, we obtain the following lemma.
\begin{lemma}
    Let $K_3(G^{\mathbf{\mu, \upsilon}}(\mathbf{X}))$ be the number of triangles in the graph $G^{\mathbf{\mu, \upsilon}}(\mathbf{X})$. The number of $k$-clique in the graph $\mathsf{G}^{\mathbf{\mu, \upsilon}}(\mathbf{X})$ is $K_3(G^{\mathbf{\mu, \upsilon}}(\mathbf{X})) \cdot (n/3)^{k - 3}$. \label{lemNumKk}
\end{lemma}

\begin{proof}
Since there is no edge between nodes in the same set, a set $S$ is a $k$-clique in $\mathsf{G}^{\mathbf{\mu, \upsilon}}(\mathbf{X})$ only if it contains one node in $\mathsf{U}$, one node in $\mathsf{Y}$, and one node in each of the set $\mathsf{W}_1, \dots, \mathsf{W}_{k - 2}$. Let $S \cap \mathsf{U} = \{\mathsf{u}_{i_S}\}$ and $S \cap \mathsf{Y} = \{\mathsf{y}_{j_S}\}$. We know that $S$ is a clique if and only if $\mu_{i_S} = 1$, $\upsilon_{j_S} = 1$, and $\mathbf{X}_{i_S, j_S} = 1$. Let $\mathcal{S}$ be the set of such $(i_S, j_S)$. We notice that the number of triangles in the graph $G^{\mathbf{\mu, \upsilon}}(\mathbf{X})$ is equal to $|\mathcal{S}| \cdot n/3$. On the other hand, the number of $k$-cliques in the graph $\mathsf{G}^{\mathbf{\mu, \upsilon}}(\mathbf{X})$ is equal to $|\mathcal{S}| \cdot (n/3)^{k - 2}$. We hence obtain the lemma statement.
\end{proof}

\subsubsection*{Edge Differentially Private Algorithm $\mathsf{C}$}

We then introduce a non-interactive $\epsilon$-edge differentially private algorithm $\mathsf{C}$ which estimates the number of $K_k$ in the graph $\mathsf{G}^{\mathbf{\mu, \upsilon}}(\mathbf{X})$. For each node \( \mathsf{v} \), let \( \mathbf{a}_\mathsf{v} \) represent the adjacency vector of \( \mathsf{v} \) in the graph $\mathsf{G}^{\mathbf{\mu, \upsilon}}(\mathbf{X})$. A non-interactive \(\epsilon\)-edge differential privacy algorithm, denoted by \(\mathsf{C}\), can be described as follows:

\begin{enumerate}
    \item Each node \( \mathsf{v} \) applies an \(\epsilon\)-differential privacy algorithm, denoted by \(\mathsf{L}_\mathsf{v}\) (referred to as a randomizer), to its adjacency vector \( \mathbf{a}_\mathsf{v} \). The node then sends the result \(\mathsf{L}_\mathsf{v}(\mathbf{a}_\mathsf{v})\) to a central server.
    \item The central server applies an aggregator function \(\mathcal{A}\) to the collected results from all randomizers and outputs \(\mathsf{A}\left( \langle \mathsf{L}_\mathsf{v}(\mathbf{a}_\mathsf{v}) \rangle_{\mathsf{v}} \right)\) as the final result.
\end{enumerate}
Let us now recall the graph \( \mathsf{G}^{\mathbf{\mu,\upsilon}}(\mathbf{X}) \). For a node \( \mathsf{u}_i \in \mathsf{U} \), we denote its adjacency vector in the graph as \( \mathbf{a}_{\mathsf{u}_i}(\mathbf{\mu, \upsilon, X}) \). Given any vectors \( \mathbf{\mu}, \mathbf{\mu}', \mathbf{\upsilon}, \mathbf{\upsilon}' \) where \( \mathbf{\mu}_i = \mathbf{\mu}'_i \), we have \( \mathbf{a}_{\mathsf{u}_i}(\mathbf{\mu, \upsilon, X}) = \mathbf{a}_{\mathsf{u}_i}(\mathbf{\mu', \upsilon', X}) \). Therefore, we can simplify the notation for \( \mathbf{a}_{\mathsf{u}_i}(\mathbf{\mu, \upsilon, X}) \) to \( \mathbf{a}_{\mathsf{u}_i}(\mathbf{\mu}_i, \mathbf{X}) \). Similarly, for each node $\mathsf{y}_i \in \mathsf{Y}$, \( \mathbf{a}_{\mathsf{y}_i}(\mathbf{\mu, \upsilon, X}) \) can be abbreviated as \( \mathbf{a}_{\mathsf{y}_i}(\mathbf{\upsilon}_i, \mathbf{X}) \), and \( \mathbf{a}_{\mathsf{w}_{i,j}}(\mathbf{\mu, \upsilon, X}) \) can be abbreviated as \( \mathbf{a}_{\mathsf{w}_{i,j}}(\mathbf{\mu}, \mathbf{\upsilon}). \) Since the computation of $\mathsf{w}_{i,j}$ do not involve any private information, we can assume that those computations are done at the central server.

\subsubsection*{Constructing an Estimator of Triangle Count $h(\mathsf{C})$ from $\mathsf{C}$}

In the next step, we define an estimator of triangle count $h(\mathsf{C})$ for the graph $G^{\mathbf{\mu, \upsilon}}(\mathbf{X})$ from the estimator $\mathsf{C}$. Recall that the set of nodes of the graph $G^{\mathbf{\mu, \upsilon}}(\mathbf{X})$ are $U, Y,$ and $W$. The construction can be done as follows:
\begin{enumerate}
    \item For each \( u_i \in U \), given \( \mathbf{\mu}_i \), the user possesses sufficient information to construct the adjacency vector of the node \( \mathsf{u}_i \in \mathsf{U} \). It can then compute \( \mathsf{L}_{\mathsf{u}_i}(\mu_i, \mathbf{X}) \) according to the construction \( \mathsf{C} \). In the construction \( h(\mathsf{C}) \), we have \( \mathcal{L}_{u_i}(\mu_i, \mathbf{X}) = \mathsf{L}_{\mathsf{u}_i}(\mu_i, \mathbf{X}) \). Similarly, the node \( y_i \in Y \) returns \( \mathcal{L}_{y_i}(\upsilon_i, \mathbf{X}) = \mathsf{L}_{\mathsf{y}_i}(\upsilon_i, \mathbf{X}) \) to the central server.
    \item The central server calculate $\mathsf{L}_{\mathsf{w}_{i,j}}(\mu, \upsilon)$ for all $i$ and $j$. The server then estimates the number of triangles as $\mathcal{A}(\langle \mathcal{L}_v(\mathbf{a}) \rangle_v) = \mathsf{A}\left( \langle \mathsf{L}_\mathsf{v}(\mathbf{a}_\mathsf{v}) \rangle_{\mathsf{v}} \right) / (n/3)^{k-3}$.
\end{enumerate}

By the construction, we obtain the following lemma.
\begin{lemma} Given $\mu, \upsilon,$ and $\mathbf{X}$. Let $\mathsf{C}$ be a non-interactive $\epsilon$-edge differentially private $K_k$ count of $\mathsf{G}_k^{\mathbf{\mu, \upsilon}}(\mathbf{X})$ with error $\mathsf{E}$. Then, $h(\mathsf{C})$ is a non-interactive $\epsilon$-edge differentially private triangle count of $G^{\mathbf{\mu, \upsilon}}(\mathbf{X})$ with error $\mathsf{E} / (n/3)^{k-3}$. \label{lem43}
\end{lemma}

\begin{proof}
The only information which $h(\mathsf{C})$ has from a user $v$ is $\mathsf{L}_\mathsf{v}(\mathbf{a}_\mathsf{v})$. Since the publication of $\mathsf{L}_\mathsf{v}(\mathbf{a}_\mathsf{v})$ is $\epsilon$-edge differentially private, we have that $h(\mathsf{C})$ is a non-interactive $\epsilon$-edge differentially private protocol. 

By Lemma \ref{lemNumKk} and the definition of $h(\mathsf{C})$, we obtain that:
\begin{eqnarray*}
    |\mathsf{C}(\mathsf{G}_k^{\mathbf{\mu, \upsilon}}(\mathbf{X})) - K_k(\mathsf{G}_k^{\mathbf{\mu, \upsilon}}(\mathbf{X}))| & = & \mathsf{E},\\
    |(n/3)^{k-3} \cdot h(\mathsf{C})(G^{\mathbf{\mu, \upsilon}}(\mathbf{X})) - (n/3)^{k-3} \cdot K_3(\mathsf{G}_k^{\mathbf{\mu, \upsilon}}(\mathbf{X}))| & = & \mathsf{E},\\ 
    |h(\mathsf{C})(G^{\mathbf{\mu, \upsilon}}(\mathbf{X})) - K_3(\mathsf{G}_k^{\mathbf{\mu, \upsilon}}(\mathbf{X}))| & = & \mathsf{E} / (n/3)^{k-3}.
\end{eqnarray*}
\end{proof}

\subsubsection*{Main Result}

We are now ready to prove our main result.

\begin{theorem} Let $\mathsf{n} = \frac{k \cdot n}{3}$ be the number of nodes in the graph $\mathsf{G}_k ^{\mathbf{\mu, \upsilon}}(\mathbf{X})$.
    There exists no non-interactive $\epsilon$-differentially private algorithm which can estimate the $K_k$ count in the graph $\mathsf{G}_k ^{\mathbf{\mu, \upsilon}}(\mathbf{X})$ with expected $\ell_2$-error in $o(\mathsf{n}^{k - 1})$.
\end{theorem}
\begin{proof}
Consider the contradictory statement that there exists a non-interactive \(\epsilon\)-differentially private algorithm capable of estimating the \(K_k\) count in the graph \(\mathsf{G}_k ^{\mathbf{\mu, \upsilon}}(\mathbf{X})\) with an expected \(\ell_2\)-error of \(o(\mathsf{n}^{k - 1})\). According to Lemma \ref{lem43}, we can construct \(h(\mathsf{C})\), which represents the triangle count in the graph \(\mathsf{G}_k ^{\mathbf{\mu, \upsilon}}(\mathbf{X})\), with an expected \(\ell_2\)-error of \(o(\mathsf{n}^{k - 1} / (n/3)^{k - 3}) = o((k \cdot n/3)^{k - 1} / (n/3)^{k - 3}) = o(n^2)\). This implies that 
$$\mathbb{E}\left[|h(\mathsf{C})(G^{\mathbf{\mu, \upsilon}}(\mathbf{X})) - K_3(\mathsf{G}_k^{\mathbf{\mu, \upsilon}}(\mathbf{X}))|^2\right] = o(n^4).$$
We can use the Jensen's inequality to show that 
$$\mathbb{E}\left[|h(\mathsf{C})(G^{\mathbf{\mu, \upsilon}}(\mathbf{X})) - K_3(\mathsf{G}_k^{\mathbf{\mu, \upsilon}}(\mathbf{X}))|\right] = o(n^2).$$
Let $\gamma = 10^{-7}$. We know that, for large $n$,  
$$\mathbb{E}\left[|h(\mathsf{C})(G^{\mathbf{\mu, \upsilon}}(\mathbf{X})) - K_3(\mathsf{G}_k^{\mathbf{\mu, \upsilon}}(\mathbf{X}))|\right] < \gamma \cdot n^2.$$
Using Markov's inequality, we obtain that
$$\mathrm{Pr}\left[|h(\mathsf{C})(G^{\mathbf{\mu, \upsilon}}(\mathbf{X})) - K_3(\mathsf{G}_k^{\mathbf{\mu, \upsilon}}(\mathbf{X}))| \geq \frac{n^2}{12}\right] < 12 \cdot \gamma.$$
Hence, 
$$\mathrm{Pr}\left[|h(\mathsf{C})(G^{\mathbf{\mu, \upsilon}}(\mathbf{X})) - K_3(\mathsf{G}_k^{\mathbf{\mu, \upsilon}}(\mathbf{X}))| \leq \frac{n^2}{12}\right] > 1 - 12 \cdot \gamma,$$
$$\mathbb{E} \left[ \left|\left\{ t \in \{1, \dots, \mathsf{N} \} : \left|h(\mathsf{C})(\mu^{(t)}, \nu^{(t)}, \mathbf{X}) - K_3\left(G^{\mu^{(t)}, \nu^{(t)}} (\mathbf{X})\right)\right| \leq \frac{n^2}{12} \right\} \right| \right] > (1 - 12 \cdot \gamma) \mathsf{N}.$$
By the Chernoff's bound, we obtain that:
$$\mathrm{Pr} \left[ \left|\left\{ t \in \{1, \dots, \mathsf{N} \} : \left|h(\mathsf{C})(\mu^{(t)}, \nu^{(t)}, \mathbf{X}) - K_3\left(G^{\mu^{(t)}, \nu^{(t)}} (\mathbf{X})\right)\right| \leq \frac{n^2}{12} \right\} \right| \leq \frac{\mathsf{N}}{5184} \right] < e^{-2((1 - 12\gamma - \frac{1}{5184}) \mathsf{N})^2 / \mathsf{N}} < 0.36.$$
Thus,
$$\mathrm{Pr} \left[ \left|\left\{ t \in \{1, \dots, \mathsf{N} \} : \left|h(\mathsf{C})(\mu^{(t)}, \nu^{(t)}, \mathbf{X}) - K_3\left(G^{\mu^{(t)}, \nu^{(t)}} (\mathbf{X})\right)\right| \leq \frac{n^2}{12} \right\} \right| \geq \frac{\mathsf{N}}{5184} \right] > 0.64.$$
Since \( h(\mathsf{C}) \) can be regarded as \( \mathcal{D}_\mathsf{C} \), this leads to a contradiction with Theorem \ref{thm41}.
\end{proof}

\end{document}